# Phase analysis and full phase control of chip-scale infrared frequency combs

Luigi Consolino[a], Francesco Cappelli[a], Pablo Cancio[a,b], Iacopo Galli[a,b], Davide Mazzotti[a,b], Saverio Bartalini[a,b], Paolo De Natale*[a]

[a]CNR-INO – Istituto Nazionale di Ottica, Largo Enrico Fermi 6, 50125 Firenze FI, Italy
& LENS – European Laboratory for Non-Linear Spectroscopy, Via Nello Carrara 1, 50019 Sesto Fiorentino FI, Italy
[b]ppqSense Srl, Via Gattinella 20, 50013 Campi Bisenzio FI, Italy

**ABSTRACT**

The road towards the realization of quantum cascade laser (QCL) frequency combs (QCL-combs) has undoubtedly attracted ubiquitous attention from the scientific community, as these devices promise to deliver all-in-one (i.e. a single, miniature, active devices) frequency comb (FC) synthesizers in a range as wide as QCL spectral coverage itself (from about 4 microns to the THz range), with the unique possibility to tailor their spectral emission by band structure engineering. For these reasons, vigorous efforts have been spent to characterize the emission of four-wave-mixing multi-frequency devices, aiming to seize their functioning mechanisms. However, up to now, all the reported studies focused on free-running QCL-combs, eluding the fundamental ingredient that turns a FC into a useful metrological tool. For the first time we have combined mode-locked multi-frequency QCL emitters with full phase stabilization and independent control of the two FC degrees of freedom. At the same time, we have introduced the Fourier transform analysis of comb emission (FACE) technique, used for measuring and simultaneously monitoring the Fourier phases of the QCL-comb modes. The demonstration of tailored-emission, miniaturized, electrically-driven, mid-infrared/THz coverage, fully-stabilized and fully-controlled QCL-combs finally enables this technology for metrological-grade applications triggering a new scientific leap affecting several fields ranging from everyday life to frontier-research.

## 1. INTRODUCTION

In the last years, environmental sensing has gained from the relentless progress of laser-based spectrometers, in terms of selectivity and sensitivity. This led to spectacular results using more and more miniaturized setups with continuous performance improvements, recently approaching the metrological-grade, with the combined use of FCs[1-3]. In parallel, sensitivity to molecular detection has dramatically improved, especially thanks to the shift of the spectral window of observation towards the mid-infrared (MIR) and THz ranges and to the introduction of novel spectroscopic techniques. The improved performance of MIR QCLs[4-8], together with the understanding of their internal dynamics[9,10] and the development of effective techniques for their frequency stabilization and linewidth narrowing[11,12] moved forward the frontiers of trace-gas detection and infrared frequency metrology[13-15].

Most recent research activities in this field aim at creating compact and integrated devices for fast and broadband spectroscopy in the MIR. Among them, QCLs generating FCs (QCL-combs) play a key role. The capability of these devices to generate FCs was demonstrated only few years ago both in the MIR and in the THz[16,17], and they are currently experiencing intense research and development[18-21]. In order to apply them for high-resolution molecular spectroscopy and metrology, a full frequency stabilization of all the emitted modes is required, together with a technique enabling to monitor the obtained degree of coherence.

*paolo.denatale@ino.cnr.it

The full phase stabilization and independent control of the two FC degrees of freedom, offset and mode spacing, has been recently demonstrated for the first time for THz QCL-combs[22]. By combining driving current modulation and radio-frequency (RF) injection a high degree of phase coherence is achieved. The obtained degree of coherence is evaluated by monitoring the Fourier phases of the FC modes with an analysis based on dual-comb multiheterodyne detection using a metrological FC as reference (LO-FC) and a subsequent Fourier transform. This technique, named FACE, has been recently introduced[23]: It enables a simultaneous retrieval of the modes phases.

## 2. DISCUSSION

**Fourier-transform analysis of comb emission (FACE)**

The core of the FACE approach is the dual-comb multiheterodyne detection. The scheme is used to compare the emission of the FC that must be controlled and characterized (sample FC) against a well-known metrological reference FC (LO-FC), obtaining a down-converted RF FC[23]. The characterization addresses a MIR and a THz QCL-comb, using two diverse difference-frequency-generated FCs[24-28] (DFG-FCs) as LO-FCs (see fig. 1).

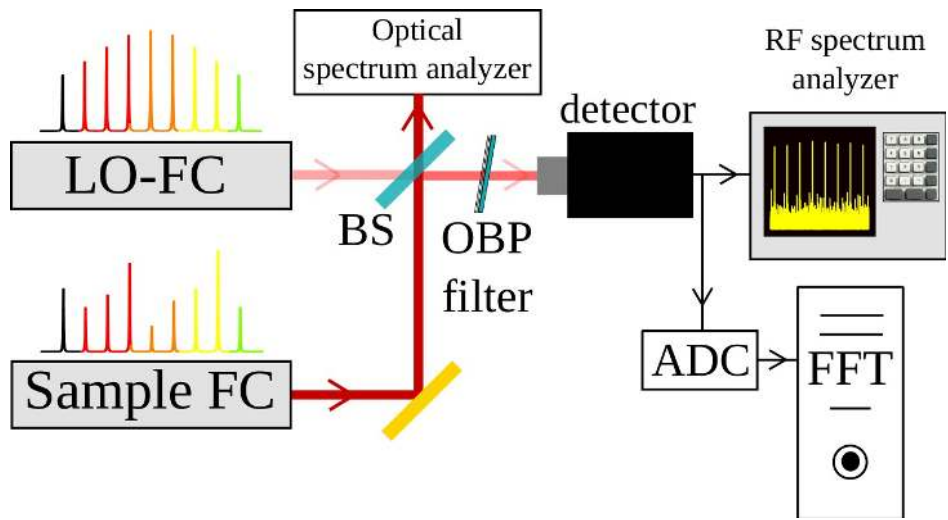

Figure 1. Dual-comb multi-heterodyne detection scheme. BS: beam splitter. OBP: optical bandpass filter. ADC: analog-to-digital converter.

The signal is acquired as time traces and then converted to the frequency domain through Fourier transformation (FFT), obtaining frequency, amplitude and phase of the sample FC modes against the ones of the LO-FC (see fig. 2). Several acquisitions have been performed, allowing an investigation of the stability of the modes phases on different time scales.

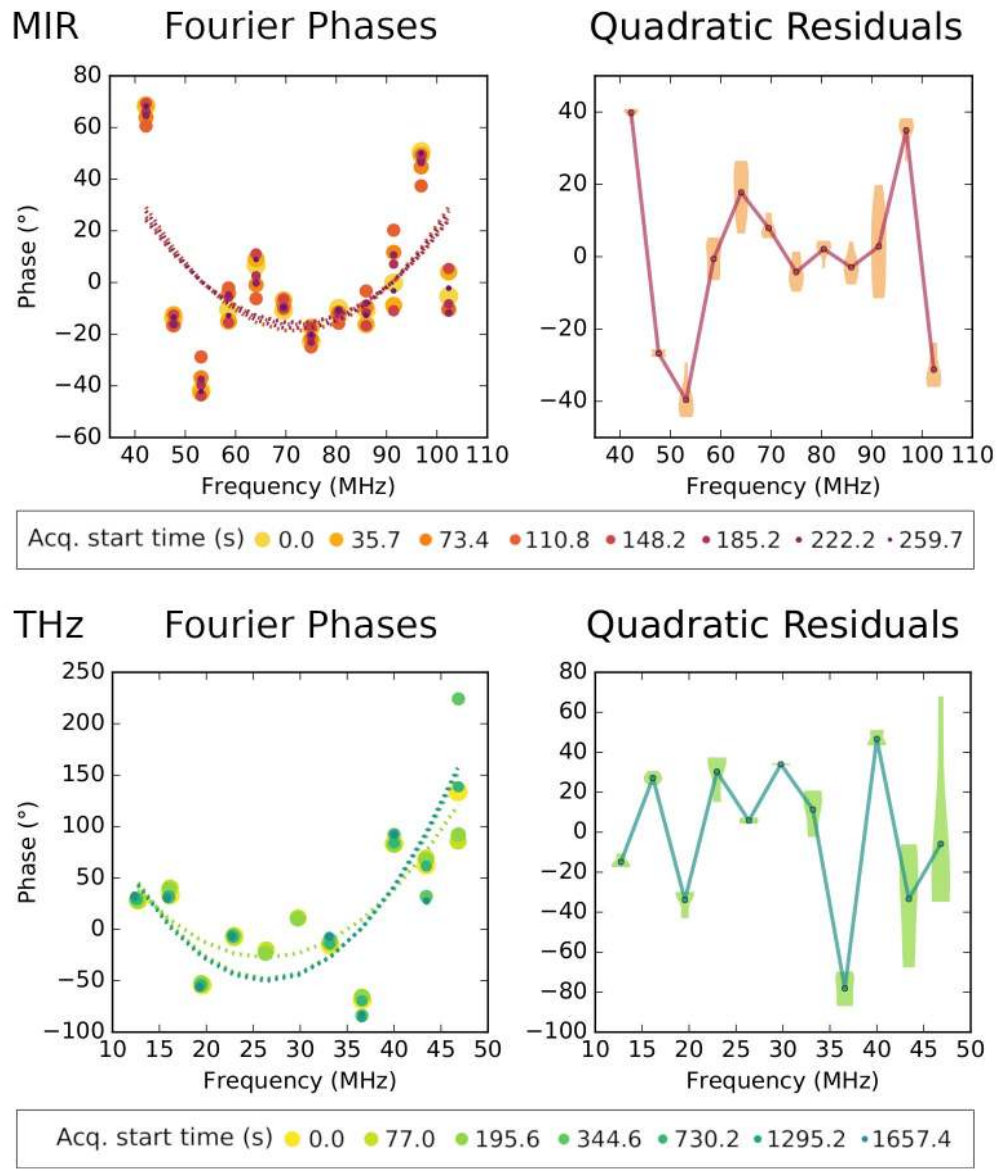

Figure 2. **Left:** Fourier phases of the MIR and the THz QCL-combs modes related to 8 (7) acquisitions. **Right:** Quadratic residuals of the phases.

For the THz QCL-comb all the emitted modes can be acquired, therefore it is possible to reconstruct the emission profile and the instantaneous frequency (fig. 3). A hybrid frequency/amplitude modulation type of FC emission is found.

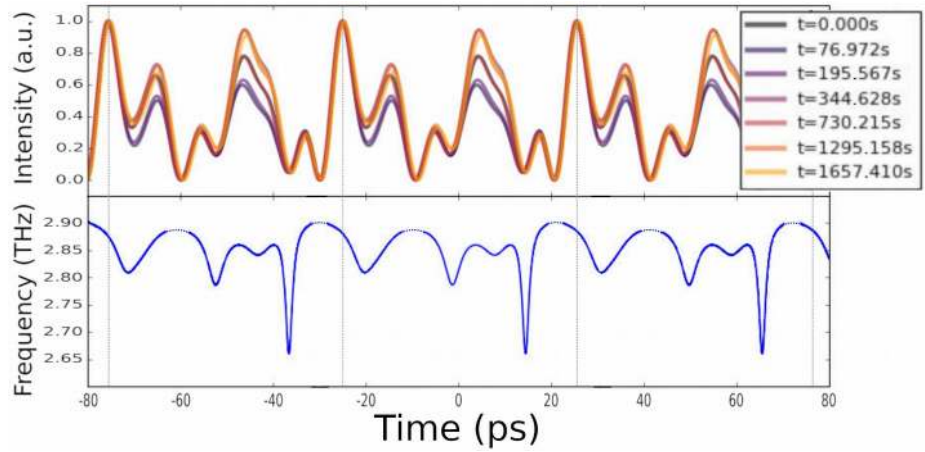

Figure 3. Reconstructed emission intensity profile and instantaneous laser frequency.

**Full phase control of FC emission**

For a THz QCL-comb it has also been possible to fully stabilize the emission by independently controlling the two degrees of freedom. The mode spacing has been stabilized by modulating the driving current at the cavity round-trip by using a referenced RF signal (RF injection), while the offset has been controlled by phase-locking one of the sample FC modes to the closest LO-FC mode[22]. In fig. 4 the obtained metrological-grade frequency stability can be clearly noticed.

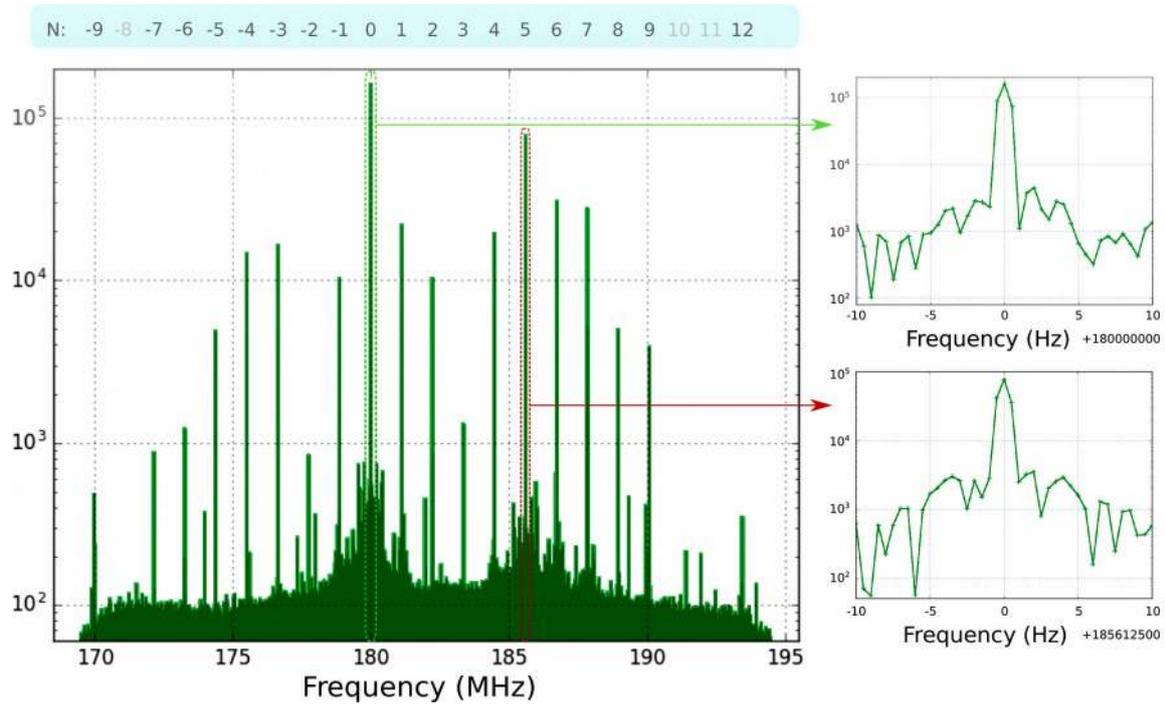

Figure 4. FFT of the multiheterodyne signal related to the fully-stabilized THz QCL-comb. All the beatnotes are resolution bandwidth (0.5 Hz) limited.

With the same actuators the independent tunability of the two FC degrees of freedom has also been demonstrated (see fig. 5).

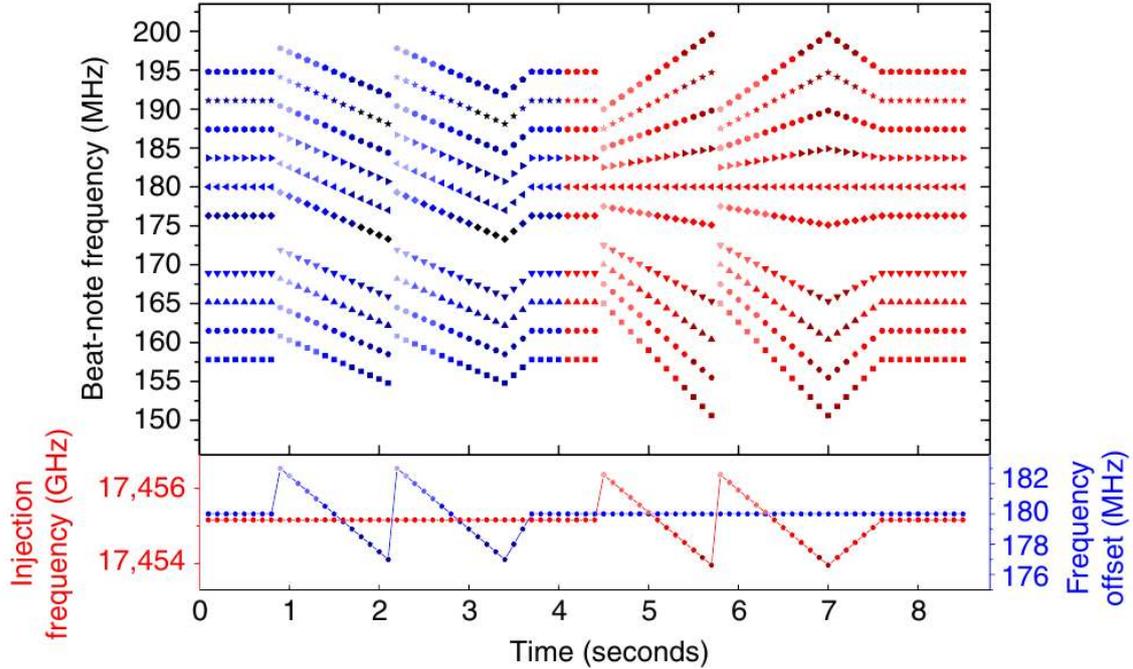

Figure 5. Demonstration of the tunability of the two FC degrees of freedom. First the offset is tuned while keeping the mode spacing constant. Then the mode spacing is tuned while keeping the offset constant.

## 3. CONCLUSIONS

The phase stability attained for the fully-stabilized QCL-comb (10° of fluctuations of the phases over several minutes of observation) conclusively proves the high coherence of the emission of these devices[23]. This evidence, together with the demonstration of the independent tunability of the two degrees of freedom[22], demonstrates that these sources are suitable for highly-demanding sensing and metrological applications.

## ACKNOWLEDGMENTS

The authors acknowledge financial support from the European Union's Horizon 2020 research and innovation programme (Laserlab-Europe Project, grant no. 654148; Qombs Project, FET Flagship on Quantum Technologies grant no. 820419) and the Italian ESFRI Roadmap (Extreme Light Infrastructure - ELI Project).

# REFERENCES

[1] T. Udem et al., "Optical frequency metrology," Nature **416**, 233-237 (2002).
[2] S. A. Diddams, "The evolving optical frequency comb," J. Opt. Soc. Am. B **27**, B51 (2010).
[3] P. Maddaloni, P. De Natale, and M. Bellini, Laser-based measurements for time and frequency domain applications: a handbook (CRC Press, Boca Raton, FL-USA, 2013).
[4] J. Faist et al., "Quantum cascade laser," Science **264**, 553-556 (1994).
[5] M. Beck et al., "Continuous wave operation of a mid-infrared semiconductor laser at room temperature," Science **295**, 301-305 (2002).
[6] Y. Bai et al., "Quantum cascade lasers that emit more light than heat," Nature Photon. **4**, 99-102 (2010).
[7] M. Razeghi et al., "Recent advances in mid infrared (3-5 µm) Quantum Cascade Lasers," Opt. Mater. Express **3**, 1872-1884 (2013).
[8] S. Riedi et al., "Broadband superluminescence, 5.9 µm to 7.2 µm, of a quantum cascade gain device," Opt. Express **23**, 7184-7189 (2015).
[9] S. Bartalini et al., "Observing the intrinsic linewidth of a quantum-cascade laser: beyond the Schawlow-Townes limit," Phys. Rev. Lett. **104**, 083904 (2010).
[10] L. Tombez et al., "Wavelength tuning and thermal dynamics of continuous-wave mid-infrared distributed feedback quantum cascade lasers," Appl. Phys. Lett. **103**, 031111 (2013).
[11] F. Cappelli et al., "Subkilohertz linewidth room-temperature mid-infrared quantum cascade laser using a molecular sub-doppler reference," Opt. Lett. **37**, 4811-4813 (2012).
[12] I. Galli et al., "Comb-assisted subkilohertz linewidth quantum cascade laser for high-precision mid-infrared spectroscopy," Appl. Phys. Lett. **102**, 121117 (2013).
[13] L. Consolino et al., "QCL-based frequency metrology from the mid-infrared to the THz range: a review," Nanophotonics **8**, 181-204 (2018).
[14] S. Borri et al., "High-precision molecular spectroscopy in the mid-infrared using quantum cascade lasers," Appl. Phys. B **125**, 18 (2019).
[15] I. Galli et al., "Absolute frequency measurements of $CO_2$ transitions at 4.3 µm with a comb-referenced quantum cascade laser," Mol. Phys. **111**, 2041-2045 (2013).
[16] A. Hugi et al., "Mid-infrared frequency comb based on a quantum cascade laser," Nature **492**, 229-233 (2012).
[17] D. Burghoff et al., "Terahertz laser frequency combs," Nat. Photon. **8**, 462-467 (2014).
[18] G. Villares et al., "Dual-comb spectroscopy based on quantum-cascade-laser frequency combs," Nat. Commun. **5**, 5192 (2014).
[19] F. Cappelli et al., "Intrinsic linewidth of quantum cascade laser frequency combs," Optica **2**, 836-840 (2015).
[20] F. Cappelli et al., "Frequency stability characterization of a quantum cascade laser frequency comb," Laser Photon. Rev. **10**, 623-630 (2016).
[21] M. Singleton et al., "Pulses from a mid-infrared quantum cascade laser frequency comb using an external compressor," J. Opt. Soc. Am. B **36**, 1676-1683 (2019).
[22] L. Consolino et al., "Fully phase-stabilized quantum cascade laser frequency comb," Nat. Commun. **10**, 2938 (2019).
[23] F. Cappelli, L. Consolino et al., "Retrieval of phase relation and emission profile of quantum cascade laser frequency combs," Nat. Photon. **13**, 562-568 (2019).
[24] I. Galli et al., "High-coherence mid-infrared frequency comb," Opt. Express **21**, 28877 (2013).
[25] I. Galli et al., "Mid-infrared frequency comb for broadband high precision and sensitivity molecular spectroscopy," Opt. Lett. **39**, 5050 (2014).
[26] G. Campo et al., "Shaping the spectrum of a down-converted mid-infrared frequency comb," J. Opt. Soc. Am. B **34**, 2287 (2017).
[27] L. Consolino et al., "Phase-locking to a free-space terahertz comb for metrological-grade terahertz lasers," Nat. Commun. **3**, 1040 (2012).
[28] M. De Regis et al., "Waveguided approach for difference frequency generation of broadly-tunable continuous-wave terahertz radiation," Appl. Sci. **8**, 2374 (2018).